%% file: paper_draft_final.tex
\documentclass[aps,prl,twocolumn,superscriptaddress,showpacs,preprintnumbers,amsmath,amssymb]{revtex4}
\usepackage{graphicx} 
\usepackage{dcolumn}  
\usepackage{float}
\usepackage{subfigure}
\usepackage{color,soul}
\usepackage[colorlinks,citecolor=blue,urlcolor=blue,bookmarks=false,hypertexnames=true]{hyperref}

\RequirePackage{xspace}

\def\qqbar  {\ensuremath{q\overline q}\xspace}

\def\pip    {\ensuremath{\pi^+}\xspace}
\def\pim    {\ensuremath{\pi^-}\xspace}
\def\pipm   {\ensuremath{\pi^{\pm}}\xspace}
\def\Kbar   {\kern 0.2em\overline{\kern -0.2em K}{}\xspace}

\def\Kz     {\ensuremath{K^0}\xspace}
\def\Kzb    {\ensuremath{\Kbar^0}\xspace}
\def\KzKzb  {\ensuremath{\Kz \kern -0.16em \Kzb}\xspace}
\def\Kp     {\ensuremath{K^+}\xspace}
\def\Km     {\ensuremath{K^-}\xspace}

\def\kpm    {\ensuremath{K^{\pm}}\xspace}
\def\KS     {\ensuremath{K^0_{\scriptscriptstyle S}}\xspace}

\def\Dbar   {\kern 0.2em\overline{\kern -0.2em D}{}\xspace}

\def\Dz     {\ensuremath{D^0}\xspace}
\def\Dzb    {\ensuremath{\Dbar^0}\xspace}
\def\DzDzb  {\ensuremath{\Dz {\kern -0.16em \Dzb}}\xspace}
\def\Dp     {\ensuremath{D^+}\xspace}
\def\Dm     {\ensuremath{D^-}\xspace}
\def\DpDm   {\ensuremath{\Dp {\kern -0.16em \Dm}}\xspace}

\def\Dstarp {\ensuremath{D^{*+}}\xspace}

\def\Bbar   {\kern 0.18em\overline{\kern -0.18em B}{}\xspace}

\def\BB     {\ensuremath{B\Bbar}\xspace}

\def\Bu     {\ensuremath{B^+}\xspace}
\def\Bub    {\ensuremath{B^-}\xspace}
\def\Bp     {\ensuremath{\Bu}\xspace}
\def\Bm     {\ensuremath{\Bub}\xspace}
\def\Bpm    {\ensuremath{B^{\pm}}\xspace}

\def\hu     {\ensuremath{h^+}\xspace}
\def\hp     {\ensuremath{\hu}\xspace}
\newcommand{\mkk}{\ensuremath{M_{K_{S}^{0}K_{S}^{0}}}\xspace}
\newcommand{\mkkp}{\ensuremath{M_{K_{S}^{0}K^{+}}}\xspace}
\mathchardef\Upsilon="7107
\def\Y#1S{\ensuremath{\Upsilon{(#1S)}}\xspace}
\def\mbc    {\mbox{$M_{\rm bc}$}\xspace}
\def\DeltaE {\mbox{$\Delta E$}\xspace}
\def\cm   {\ensuremath{{\rm \,cm}}\xspace}
\def\invfb{\ensuremath{\mbox{\,fb}^{-1}}\xspace}

\def\to{\ensuremath{\rightarrow}\xspace}
\def\CP {\ensuremath{C\!P}\xspace}
\def\ACP{{\ensuremath{\mathcal{A}_{\CP}}\xspace}}
\def\etal  {{\it et~al.}}
\def\nb    {\ensuremath{C_\mathrm{{N\!B}}}\xspace}
\def\nbprim{\ensuremath{C'_\mathrm{{N\!B}}}\xspace}
\def\nbmin {\ensuremath{C_{\mathrm{N\!B},{\rm min}}}\xspace}
\def\nbmax {\ensuremath{C_{\mathrm{N\!B},{\rm max}}}\xspace}
\newcommand{\stat}{\ensuremath{\mathrm{(stat)}}\xspace}
\newcommand{\syst}{\ensuremath{\mathrm{(syst)}}\xspace}
\newcommand{\gev}{\ensuremath{\mathrm{\,Ge\kern -0.1em V}}\xspace}
\newcommand{\mev}{\ensuremath{\mathrm{\,Me\kern -0.1em V}}\xspace}
\newcommand{\gevc}{\ensuremath{{\mathrm{\,Ge\kern -0.1em V\!/}c}}\xspace}
\newcommand{\mevc}{\ensuremath{{\mathrm{\,Me\kern -0.1em V\!/}c}}\xspace}
\newcommand{\gevcc}{\ensuremath{{\mathrm{\,Ge\kern -0.1em V\!/}c^2}}\xspace}
\newcommand{\mevcc}{\ensuremath{{\mathrm{\,Me\kern -0.1em V\!/}c^2}}\xspace}

\begin{document}

\preprint{\vbox{ \hbox{   }
			\hbox{Belle Preprint {\it 2018-26}}
			\hbox{KEK Preprint {\it 2018-83}}
}}

\title{\quad\\[1.0cm] Measurements of branching fraction and direct {\boldmath $\ensuremath{C\!P}$} asymmetry in {\boldmath $\Bpm\to \KS\KS \kpm$} and a search for {\boldmath $\Bpm\to \KS\KS \pipm$}}



\input{author-pub530}


\begin{abstract}
We study charmless hadronic decays of charged $B$ mesons to the
final states $\KS\KS\kpm$ and $\KS\KS\pipm$ using a 
$711\invfb$ data sample that contains $772\times 10^6$ $\BB$ pairs, and was 
collected at the $\Y4S$ resonance with the Belle detector at the KEKB 
asymmetric-energy $e^+e^-$ collider.  For $B^{\pm} \to \KS\KS\kpm$, the measured branching fraction and direct $\CP$ asymmetry are $[10.42\pm0.43\stat\pm 0.22\syst]\times10^{-6}$ 
and [$+1.6\pm3.9\stat\pm 0.9\syst$]\%, respectively. In the absence of a statistically significant signal for $B^{\pm}\to \KS\KS\pipm$, we obtain a 90\% confidence-level upper limit on its branching fraction as $8.7 \times10^{-7}$.
\end{abstract}


\maketitle

\tighten

{\renewcommand{\thefootnote}{\fnsymbol{footnote}}}
\setcounter{footnote}{0}

\begin{figure}[!htb]
\begin{center}
\includegraphics[width=.23\textwidth]{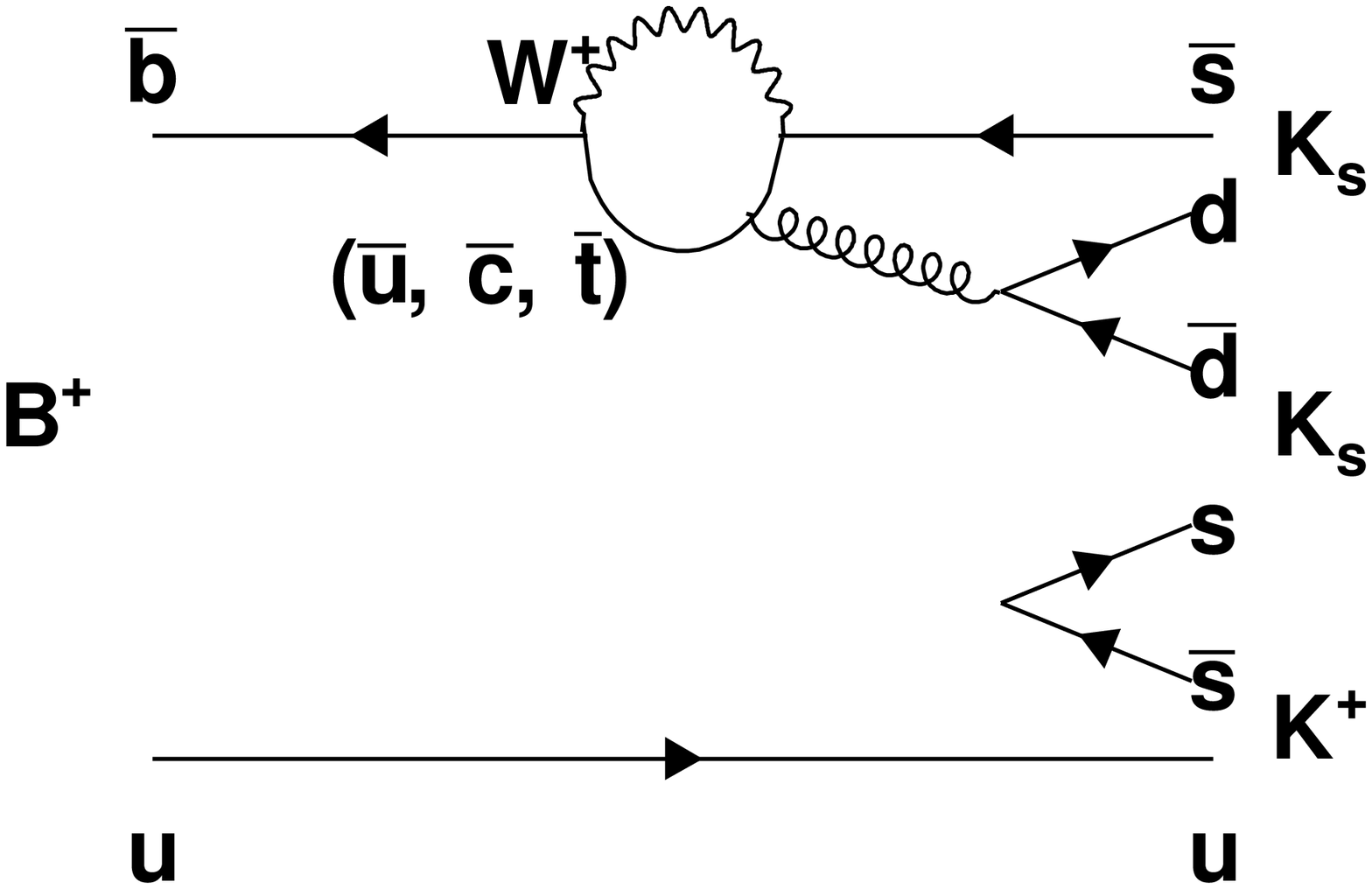} 
\includegraphics[width=.23\textwidth]{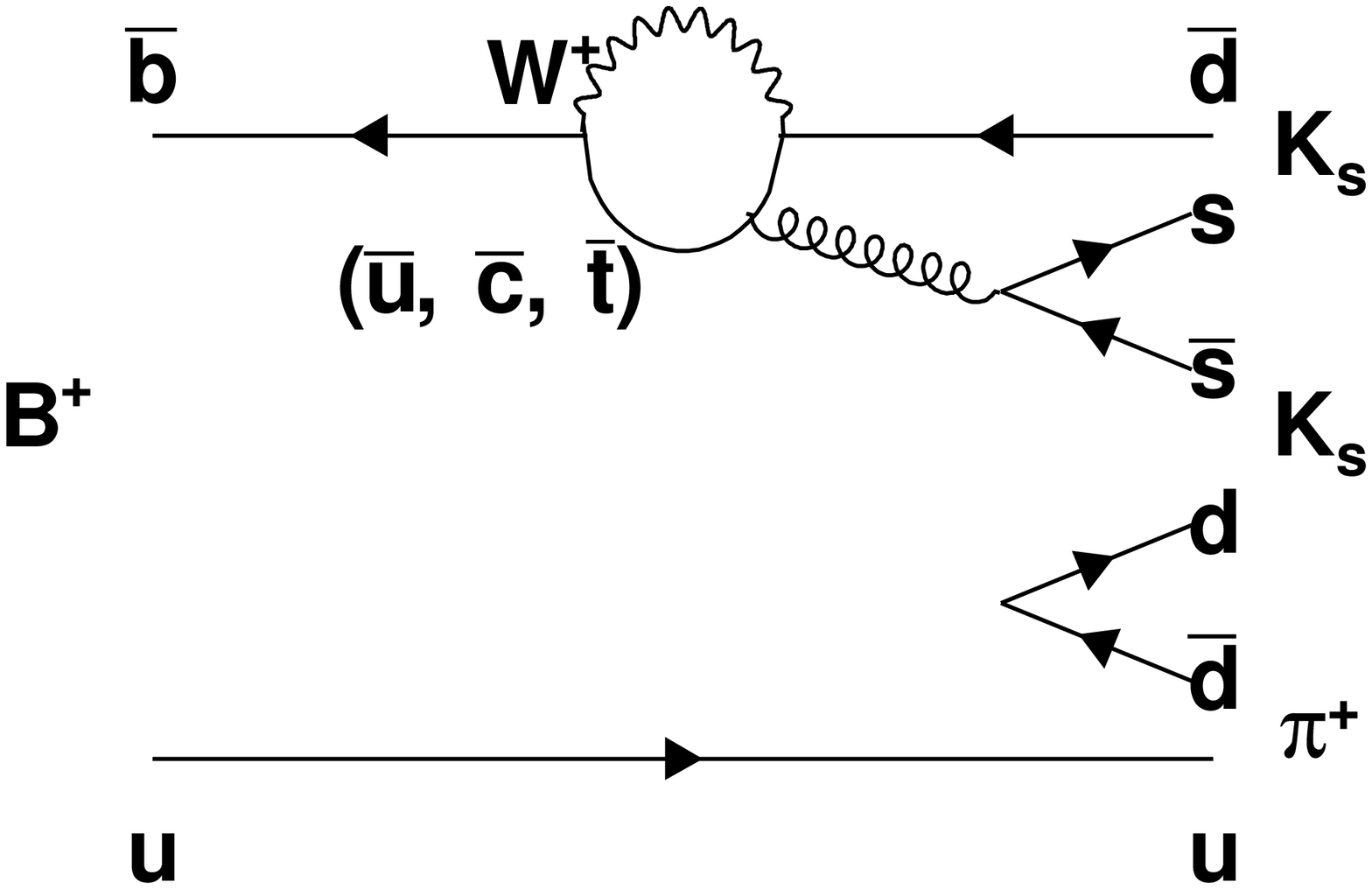} 
\end{center}
\caption{Feynman diagrams of the dominant amplitudes that contribute to the
 decays $B^{\pm}\to\KS\KS\kpm$ (left) and $B^{\pm}\to\KS\KS\pipm$ (right).}
\label{fig:Fey}
\end{figure}

Charged $B$-meson decays to the three-body charmless hadronic 
final states $\KS\KS\kpm$ and $\KS\KS\pipm$ mainly proceed 
via $b\to s$ and $b\to d$ loop transitions, respectively. Figure~\ref{fig:Fey} shows 
Feynman diagrams of the dominant amplitudes that contribute to these decays. 
These flavor changing neutral current transitions, 
being suppressed in the standard model (SM), are interesting as they could be sensitive to possible non-SM contributions~\cite{PAP:ref}. 

Further motivation, especially to study the contributions of various quasi-two-body resonances to  inclusive $\CP$ asymmetry, comes from the recent results on  $B^{\pm}\to\Kp\Km K^{\pm}$, $\Kp\Km \pi^{\pm}$ and other such three-body decays~\cite{LHCb:paper1,LHCb:paper2,Chialing:paper}. LHCb has found large asymmetries localized in phase space in $B^{\pm}\to\Kp\Km \pi^{\pm}$ decays~\cite{LHCb:paper2}. Recently, Belle has also reported strong evidence for large $\CP$ asymmetry at the low $K^{+}K^{-}$ invariant mass region of $B^{\pm}\to\Kp\Km \pi^{\pm}$~\cite{Chialing:paper}. The fact that the $K\overline{K}$ system of  $B^{\pm} \to \KS\KS h^{\pm} (h=K,\pi)$, in contrast to that of $B^{\pm} \to K^{+} K^{-} h^{\pm}$, cannot form a vector resonance (Bose symmetry) may shed light on the source of large $\CP$ violation in the latter decays.

The three-body decay $\Bp\to\KS\KS\Kp$~\cite{charge:ref1} was observed by Belle~\cite{Belle:paper1} and subsequently studied  by BaBar~\cite{BaBar:paper1}. 
Belle measured the decay branching fraction as 
$(13.4\pm1.9\pm1.5)\times10^{-6}$ based on a data sample of $70\invfb$~\cite{Belle:paper1}, and BaBar reported a branching fraction of $(10.6\pm0.5\pm0.3)\times10^{-6}$ and a $\CP$ 
asymmetry of $(+4_{-5}^{+4}\pm2)\%$ using $426\invfb$ of data~\cite{BaBar:paper1}. The quoted uncertainties are statistical and systematic, respectively. 

The decay $\Bp\to\KS\KS\pip$ is suppressed by the squared ratio of CKM matrix~\cite{CKM:ckm} elements $|V_{td}/V_{ts}|^2 (= 0.046)$ with respect to $\Bp\to\KS\KS\Kp$, and has not yet been observed. The most restrictive limit at $90\%$ confidence level on its branching fraction, 
${\cal B}(\Bp\to\KS\KS\pip)<5.1\times 10^{-7}$, comes from BaBar~\cite{BaBar:paper2}. 

We present an improved measurement of the branching fraction and direct $\CP$ asymmetry  of the decay $\Bp\to\KS\KS \Kp$ as well as a search for $\Bp\to\KS\KS\pip$  using a data sample of $711\invfb$, which contains $772\times 10^6$  $\BB$ pairs and was recorded near the $\Y4S$ resonance with the Belle
detector~\cite{Belle} at the KEKB $e^+e^-$ collider~\cite{KEKB}. The direct $\CP$ asymmetry is defined as
\begin{equation}
\ACP = \dfrac{N (\Bm\to\KS\KS h^{-}) - N(\Bp\to\KS\KS h^{+})}{N(\Bm\to\KS\KS h^{-}) + N (\Bp\to\KS\KS h^{+})}, 
\end{equation}
where $N$ is the obtained signal yield for the corresponding mode. The detector components relevant for our  study are a silicon vertex detector (SVD), a $50$-layer central drift chamber (CDC), an array of aerogel threshold Cherenkov counters (ACC), and a barrel-like arrangement of time-of-flight scintillation counters (TOF); all located inside a $1.5$\,T solenoidal magnetic field.

To reconstruct $\Bp\to\KS\KS\hp$ candidates, we begin by identifying charged kaons and pions. A kaon or pion candidate track must have a minimum transverse momentum of $100\mevc$ in the lab frame, and a distance of closest approach with respect to the interaction point (IP) 
of less than $0.2\cm$ in the transverse $r$--$\phi$ plane and less than 
$5.0\cm$ along the $z$ axis. Here, the $z$ axis is defined opposite the $e^+$ beam. Charged tracks are identified as kaons or pions based on a likelihood ratio ${\cal R}_{K/\pi}={{\cal L}_K}/({\cal L}_K+{\cal L}_\pi)$, where ${\cal L}_K$ and ${\cal L}_\pi$ are the individual likelihoods 
for kaons and pions, respectively, calculated with information from the CDC, ACC and TOF. Tracks with  
${\cal R}_{K/\pi}>0.6$ are identified as kaons while those with ${\cal R}_{K/\pi}< 0.4$ are identified as pions. The efficiency for kaon (pion) identification is $86\%$ ($91\%$) with a pion (kaon) misidentification rate of $9\%$ ($14\%$).

The $\KS$ candidates are reconstructed from pairs of oppositely charged 
tracks, both assumed to be pions, and are further subject to a selection~\cite{nisKs} based on a neural network~\cite{neurobayes}. The network uses the following input variables: the 
$\KS$ momentum in the lab frame; the distance along the $z$ axis 
between the two track helices at their closest approach; the $\KS$ flight length 
in the $r$--$\phi$ plane; the angle between the $\KS$ momentum and the vector 
joining the IP to the $\KS$ decay vertex; the angle between the pion 
momentum and the lab frame direction in the $\KS$ rest frame; the 
distances of closest approach in the $r$--$\phi$ plane between the IP and the two 
pion helices; the number of hits in the CDC for each pion track; and the presence/absence of hits in the SVD for each pion track. We require that the reconstructed invariant mass be 
between $491$ and $505 \mevcc$, corresponding to $\pm 3\sigma$ 
around the nominal $\KS$ mass~\cite{PDG} with $\sigma$ denoting the experimental resolution. 

We identify $B$ meson candidates using two kinematic variables:
the beam-energy constrained mass, $\mbc=\sqrt{E^2_{\rm beam}/c^{4}-\left|\sum_{i}
\vec{p}_i/c\right|^2}$, and the energy difference, $\DeltaE=\sum_{i}E_{i}-
E_{\rm beam}$, where $E_{\rm beam}$ is the beam energy, and $\vec{p}_i$
and $E_i$ are the momentum and energy of the $i$-th
daughter of the reconstructed $B$ candidate; all calculated in the center-of-mass (CM) frame.
For each $B$ candidate, we perform a fit constraining its daughters to come from a common vertex, whose position is consistent with the IP profile. Events with $5.271 \gevcc<\mbc<5.287\gevcc$ and $-0.10\gev<\DeltaE<0.15\gev$ 
are retained for further analysis. The $\mbc$ requirement corresponds approximately to a
$\pm 3\sigma$ window around the nominal $\Bp$ mass~\cite{PDG}. We apply a 
looser ($-6\sigma$,\,$+9\sigma$) requirement on $\DeltaE$ as it is later used to extract the signal yield. 

The average number of $B$ candidates per event is $1.1$ ($1.5$) for $\Bp\to\KS\KS\Kp$ 
($\KS\KS\pip$). In case of multiple candidates, we choose the one with the minimum $\chi^2$ value for the aforementioned vertex fit. This criterion selects the correct $B$-meson candidate in
75\% and 63\% of Monte Carlo (MC) events having more than one candidate in $\Bp\to\KS\KS\Kp$ and $\Bp\to\KS\KS\pip$, respectively.

The dominant background arises from the $e^+e^-\to\qqbar$ ($q=u,d,s,c$)
continuum process. We use observables based on
event topology to suppress it. The event shape in the CM frame is
expected to be spherical for $\BB$ events, whereas continuum
events are jetlike. We employ a neural network to separate signal from background
using the following six input variables: a Fisher discriminant formed from
$16$ modified Fox-Wolfram moments~\cite{KSFW}; the cosine of the angle
between the $B$ momentum and the $z$ axis; the cosine of the angle between
the $B$ thrust and the $z$ axis; the cosine of the angle between the
thrust axis of the $B$ candidate and that of the rest of the event; the
ratio of the second to the zeroth order Fox-Wolfram moments; and the vertex separation along the
$z$ axis between the $B$ candidate and the remaining tracks. The first five quantities are calculated in the CM frame. The neural network training is performed with simulated signal and
$\qqbar$ events. Signal and background samples are
generated with the {\textsc EvtGen} program~\cite{evtgen}; for signal we assume a uniform decay in phase space. A GEANT-based~\cite{Geant} simulation is used to model the detector response.

We require the neural network output ($\nb$)
to be greater than $-0.2$ to substantially reduce the continuum
background. For both decays, the relative signal efficiency due to this requirement is
approximately $91\%$ and the achieved continuum suppression is
close to $84\%$. The remainder of the $\nb$ distribution strongly peaks
near $1.0$ for signal, making it challenging to model it analytically. However, its transformed variable
\begin{equation}
\nbprim=\ln\left[\frac{\nb-\nbmin}{\nbmax-\nb}\right], 
\end{equation}
where $\nbmin=-0.2$ and $\nbmax\simeq 1.0$, can be parametrized by one or more Gaussian functions. We use $\nbprim$ as a fit variable along with $\DeltaE$.

The background due to charmed $B$ decays, mediated via the dominant $b\to c$ transition, 
is studied with an MC sample. The  resulting $\DeltaE$ and $\mbc$ distributions are found to peak in the signal region for both $\Bp\to\KS\KS \Kp$ and  $\Bp\to\KS\KS \pip$ decays. For $\Bp\to\KS\KS\Kp$, the peaking background predominantly stems from $\Bp\to\Dz \Kp$ with $\Dz\to \KS\KS$ and $\Bp\to\chi_{c0}({\rm{1P}}) \Kp$ with $\chi_{c0}({\rm{1P}})\to \KS\KS$. To suppress these backgrounds, we exclude candidates for which $\mkk$ lies in the range $[1.85,1.88]\gevcc$ or $[3.38,3.45]\gevcc$, corresponding to a $\pm 3\sigma$ window around the nominal $D^0$ or $\chi_{c0}(\rm{1P})$ mass~\cite{PDG}, respectively. In case of $\Bp\to\KS\KS\pip$, the peaking background largely arises from $\Bp\to\Dz \pip$ with $\Dz\to \KS\KS$. To suppress it, we  exclude candidates for which $\mkk$ lies in the aforementioned $D^{0}$ mass window. 

A few background modes contribute in the $\mbc$ signal
region, but having their $\DeltaE$ peak shifted from zero to the positive 
side for $\Bp\to\KS\KS\Kp$ or to the negative side for $\Bp\to\KS\KS\pip$. To identify these so-called ``feed-across'' backgrounds, mostly arising due to $K$--$\pi$ 
misidentification, we use a $\BB$ MC sample in which one of 
the $B$ mesons decays via $b\to u,d,s$ transitions, along with the charmed $\BB$ sample. For $\Bp\to\KS\KS\pip$, the feed-across background includes contributions from $\Bp\to\KS\KS\Kp$ as well as $\Bp\to\Dz \Kp$ and $\Bp\to\chi_{c0}({\rm{1P}}) \Kp$ that survive the $D^{0}$ and $\chi_{c0}(\rm{1P})$ vetoes. For $\Bp\to\KS\KS\Kp$, it comes entirely from $\Bp\to\KS\KS\pip$. All other events coming from neither the signal, continuum, nor the feed-across components form the so-called ``combinatorial''  $\BB$ background.

After all selection requirements, the efficiencies for correctly reconstructed signal events are 24\% for $\Bp\to\KS\KS\Kp$ and 26\% for $\Bp\to\KS\KS\pip$. The fractions of misreconstructed signal events for which one of the daughter particles comes from the other $B$-meson decay are  0.5\% for $\Bp\to\KS\KS\Kp$ and 1.1\% for $\Bp\to\KS\KS\pip$. We consider these events as part of the signal.

The signal yield and $\ACP$ are obtained with an unbinned extended maximum likelihood fit to the two-dimensional distribution of $\DeltaE$ and $\nbprim$. The extended likelihood function is
\begin{equation}
{\cal L} = \dfrac{\mathrm{e}^{-\sum_{j} n_{j}}}{N!} \prod_{i} \Big[\sum_{j} n_{j} \mathcal{P}_{j}^{i} \Big],
\end{equation}
where
\begin{equation}
{\cal P}_{j}^{i}\equiv\dfrac{1}{2} (1-q^{i}{\ensuremath{\mathcal{A}_{\CP ,j}}})\times {\cal P}_j(\DeltaE^{\,i})\times{\cal P}_j(\nb'^{\,i}).
\end{equation}
Here, $N$ is the total number of events, $i$ is the event index, and $n_j$ is the yield of the event category $j$ ($j\equiv$ signal, $\qqbar$, combinatorial, and feed-across). ${\cal P}_{j}$ and ${\ensuremath{\mathcal{A}_{\CP ,j}}}$ are the probability density function (PDF) and direct $\CP$ asymmetry corresponding to the category $j$, and $q^{i}$ is the electric charge of the $B$ candidate in event $i$. As the correlation between $\DeltaE$ and $\nbprim$ is small (the linear correlation coefficient ranges from 0.5\% to 7.0\%), the product of two individual PDFs is a good approximation for the total PDF. We apply a tight requirement on $\mbc$ instead of including it as a fit variable since it exhibits a large correlation with $\DeltaE$ for the signal and feed-across background. We choose $\DeltaE$ over $\mbc$ in the fit because the former is a better variable to distinguish signal from feed-across background. To account for  crossfeed  between the two channels, they are fitted simultaneously, with the $\Bp \to\KS\KS \Kp$  branching fraction in the correctly reconstructed sample determining the normalization of the crossfeed in the $\Bp \to\KS\KS \pip$ fit region, and vice versa. 
\begin{table}[htb]
\centering
\caption{List of PDFs used to model the $\DeltaE$ and $\nbprim$
distributions for various event categories for $\Bp\to\KS\KS \Kp$. G, AG, 
and Poly1 denote Gaussian, asymmetric Gaussian, and first-order polynomial, respectively.\\}
\label{tab:pdf-shape}
\begin{tabular}{lcccc}
\hline\hline
Event category & & $\DeltaE$ & & $\nbprim$ \\
\hline
Signal & & 3\,G & & G+AG \\
Continuum $\qqbar$ & & Poly1 & & 2\,G \\
Combinatorial $\BB$ & & Poly1 & & 2\,G \\
Feed-across & & G+Poly1 & & G \\
\hline\hline
\end{tabular}
\end{table} 

Table~\ref{tab:pdf-shape} lists the PDFs used to model the $\DeltaE$ and
$\nbprim$ distributions for various event categories for $\Bp\to\KS\KS \Kp$. 
For $\Bp\to\KS\KS \pip$, we use the same PDF shapes except for the feed-across
background component, where we add an asymmetric 
Gaussian function to the PDFs in Table~\ref{tab:pdf-shape} to accurately describe
$\DeltaE$ and $\nbprim$ distributions. The free parameters in the fit are the continuum background yields and the branching fractions of $\Bp\to\KS\KS \Kp$ and $\Bp\to\KS\KS \pip$,  and the signal $\ACP$ for $\Bp\to\KS\KS \Kp$. In addition, the following PDF shape parameters of the continuum background are floated in the fit for both $\Bp\to\KS\KS \Kp$ and $\KS\KS\pip$: the slope of the first-order polynomial used for $\DeltaE$ and the mean and width of the dominant Gaussian component used to model $\nbprim$.  The combinatorial $\BB$ yields are fixed to the MC values due to their correlation with the continuum yields. This is because $\nbprim$ is the only variable that offers some discrimination between the two background categories. To improve the overall fit stability, $\mathcal{A}_{\CP}$ for all components but for the $\Bp \to\KS\KS \Kp$ signal are fixed to zero. The other PDF shape parameters for signal and background
components are fixed to the corresponding MC expectations for both decays. We correct the signal $\DeltaE$ and $\nbprim$ PDF shapes for possible data-MC differences, according to the values 
obtained with a control sample of $\Bp\to\Dbar^0\pip$ with $\Dbar^0 \to \KS\pip\pim$. 
The same correction factors are also applied for the feed-across 
background component of $\Bp\to\KS\KS\pip$.

\begin{figure}
\includegraphics[width=0.492\columnwidth]{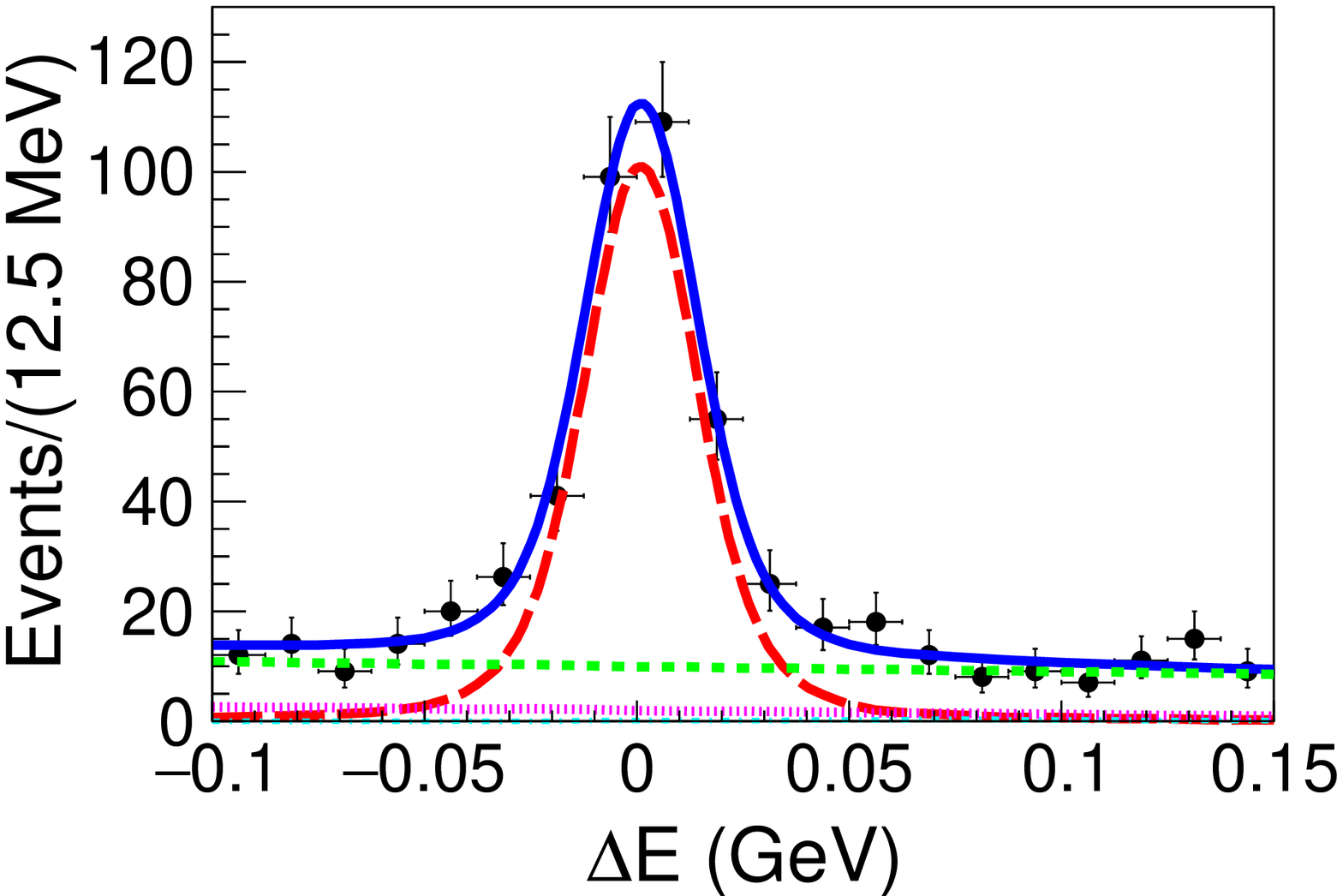}
\includegraphics[width=0.492\columnwidth]{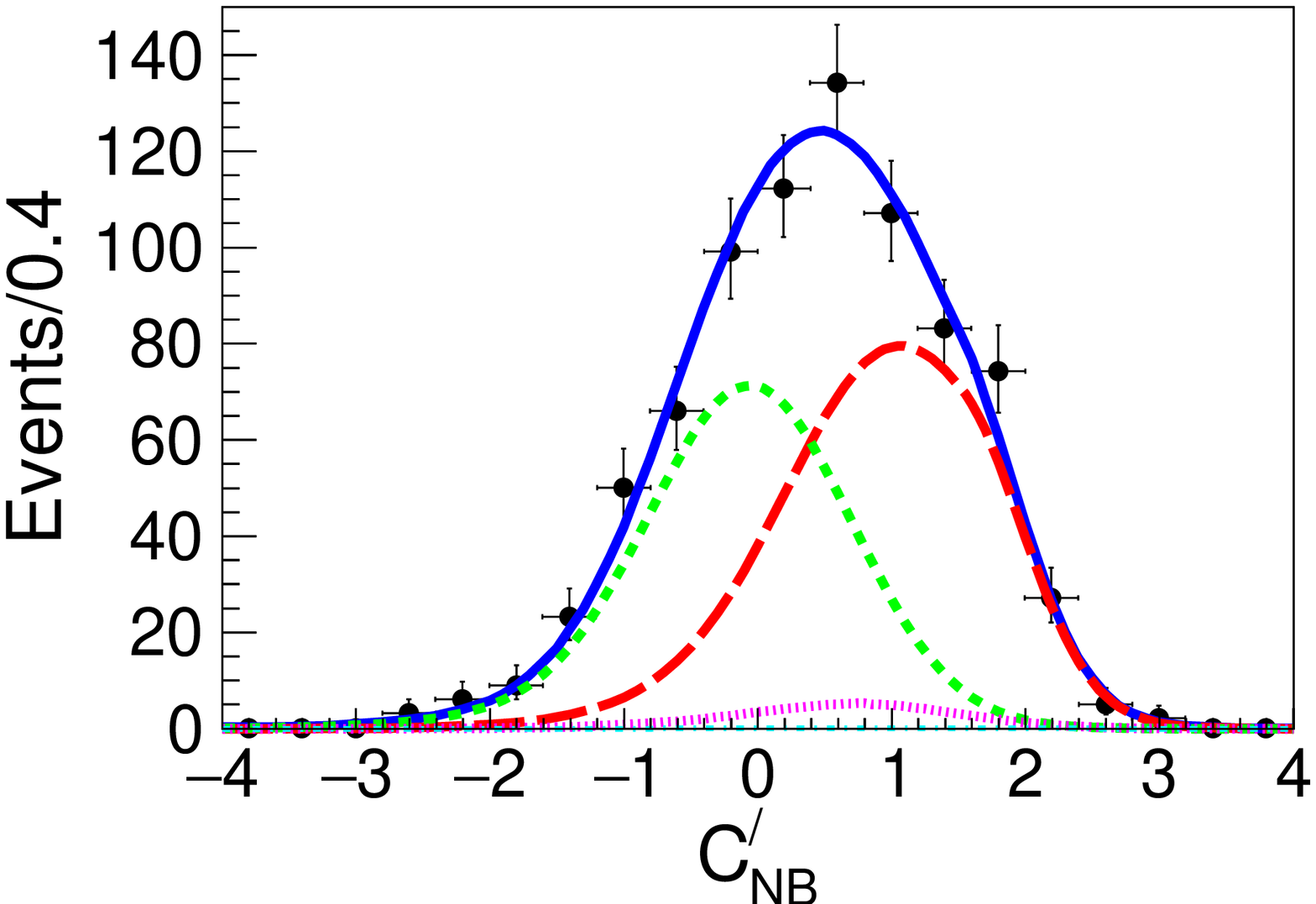}\\
\vspace{0.03 in}
  a) $B^{+}\to\KS\KS K^{+}$  \\
\vspace{0.05 in}
\includegraphics[width=0.492\columnwidth]{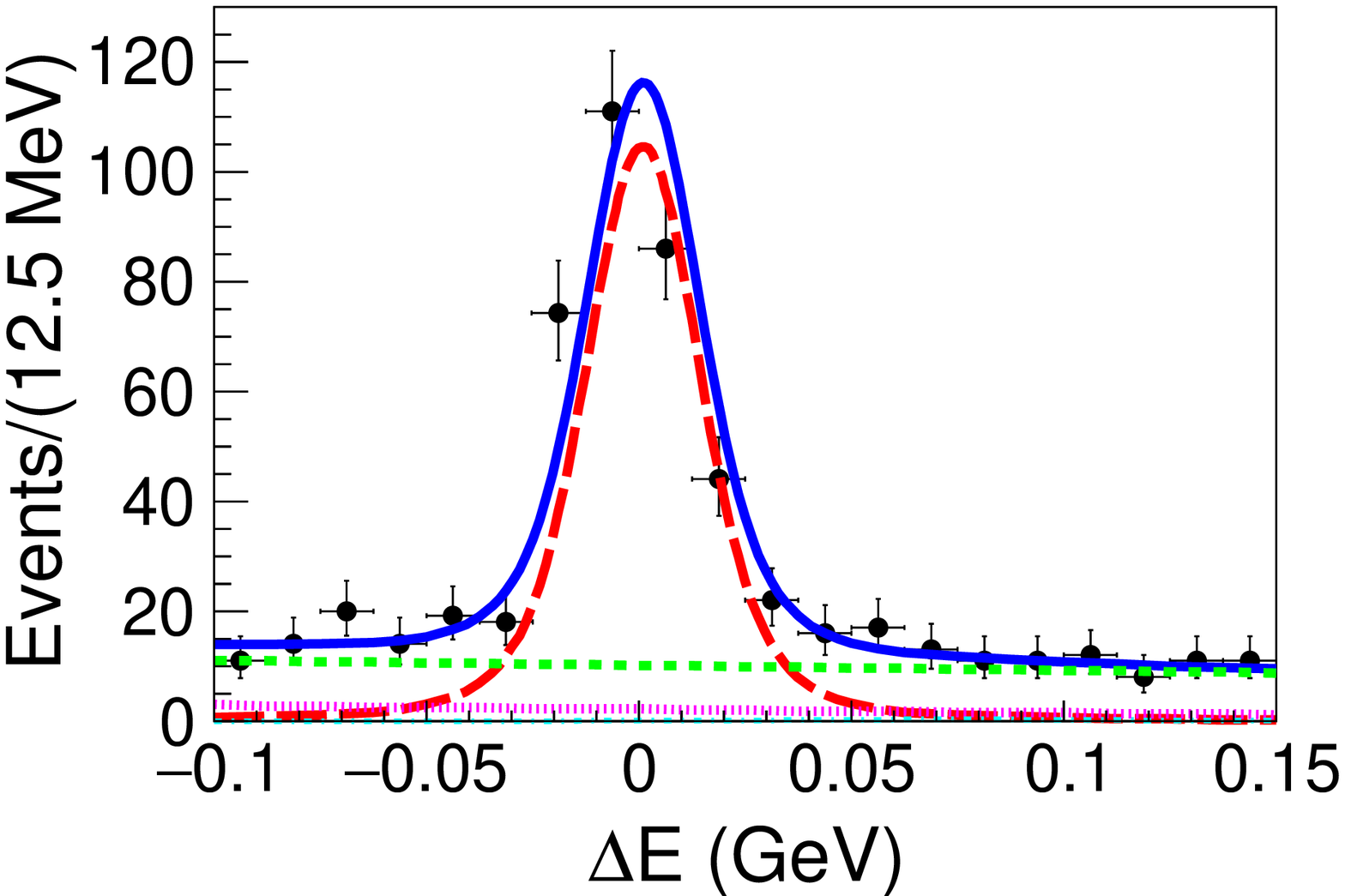}
\includegraphics[width=0.492\columnwidth]{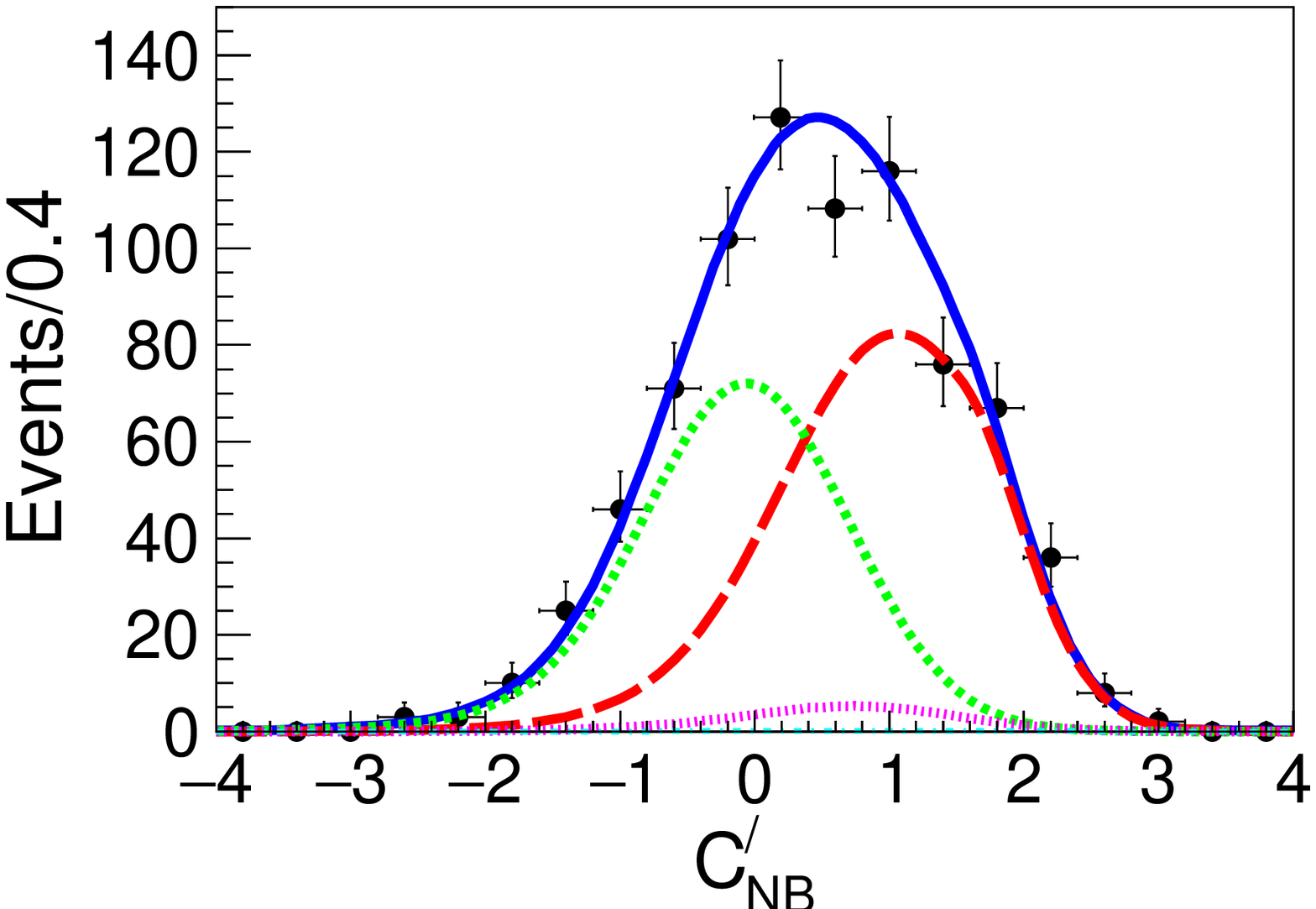}\\
\vspace{0.03 in}
  b) $B^{-}\to\KS\KS K^{-}$ \\
 \vspace{0.05 in}
\includegraphics[width=0.492\columnwidth]{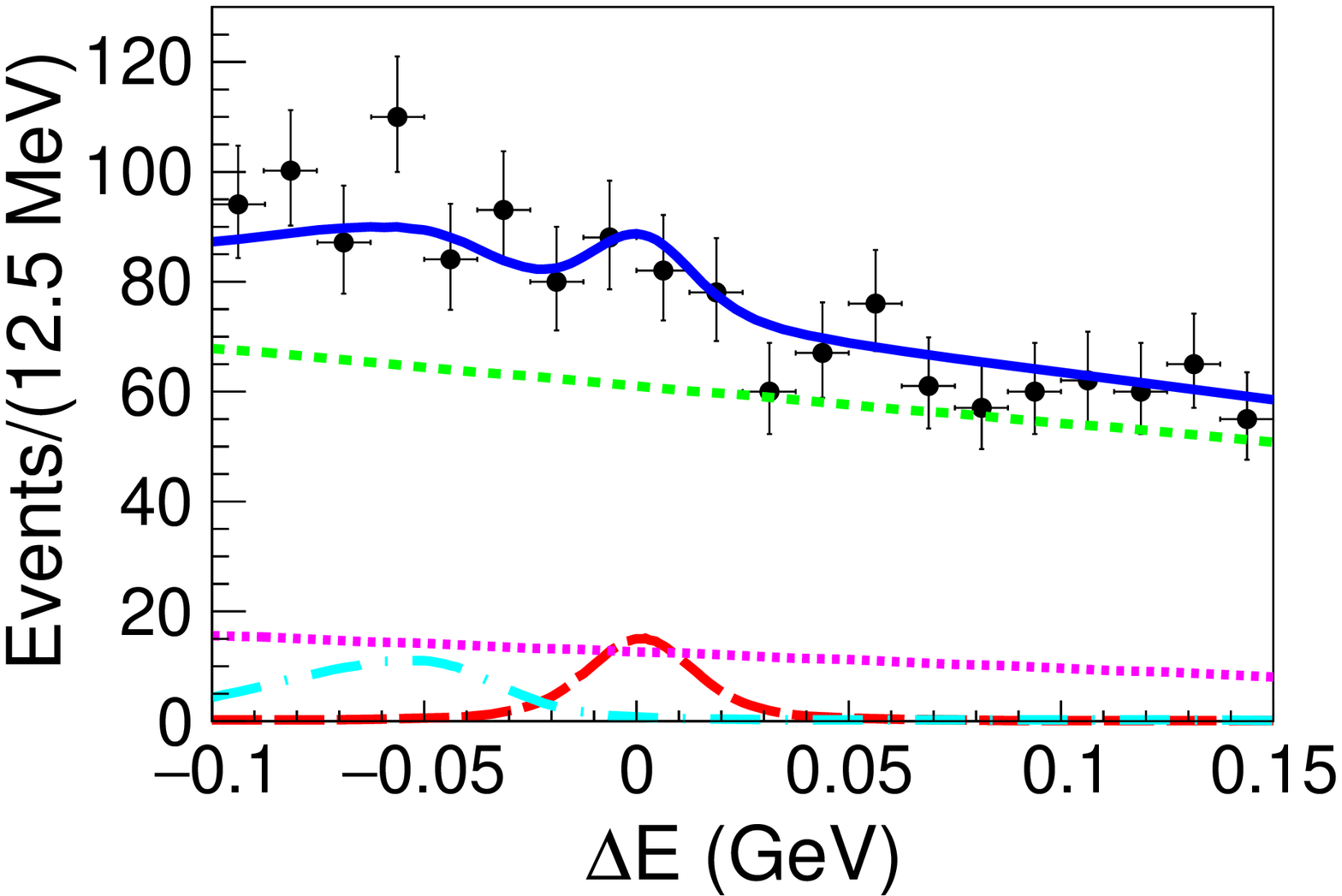}  
\includegraphics[width=0.489\columnwidth]{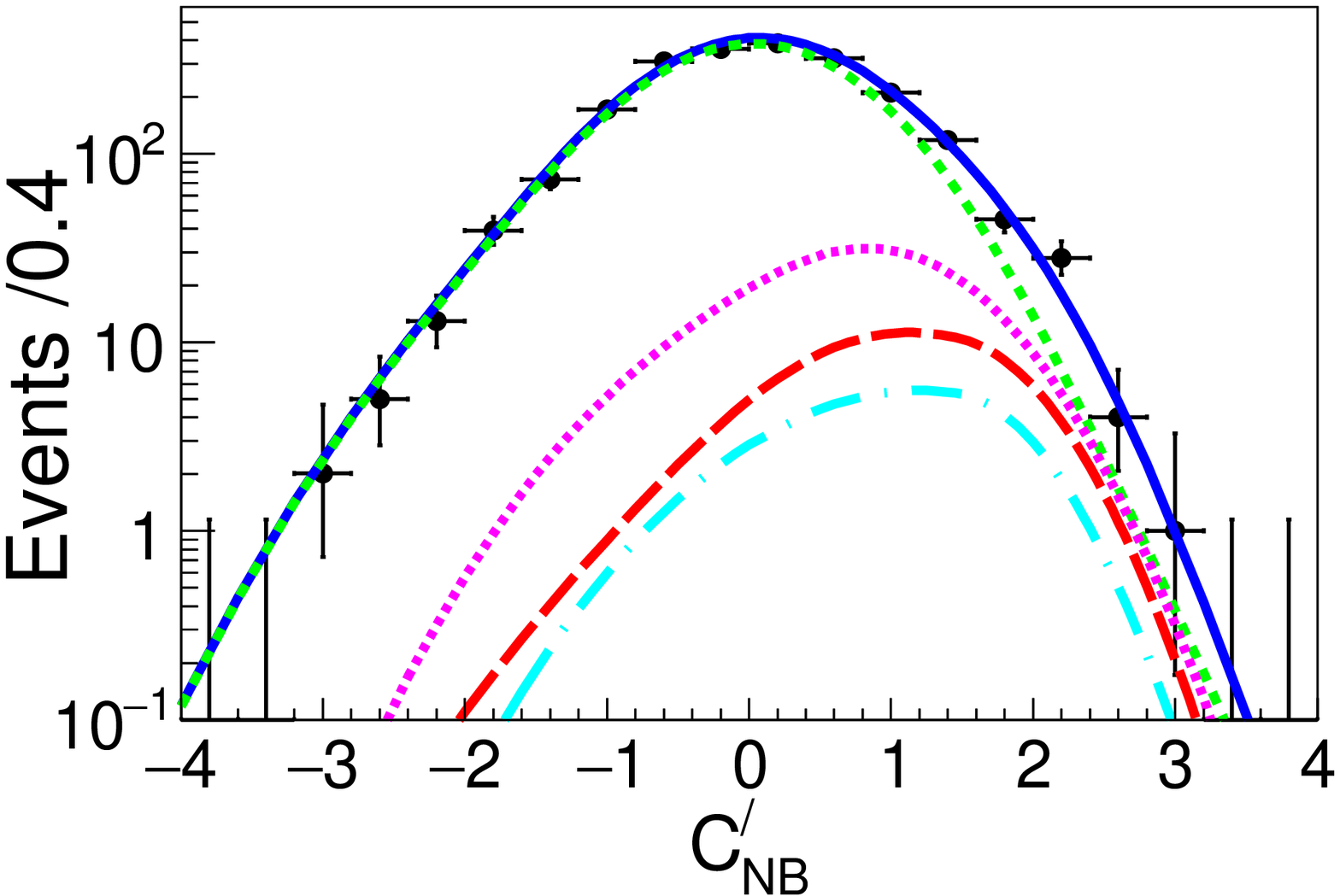}\\
\vspace{0.03 in}
   c) $B^{\pm}\to\KS\KS \pi^{\pm}$  \\
\caption{(color online). Projections of the two-dimensional simultaneous fit to $\DeltaE$ for $\nbprim>0.0$ and $\nbprim$ for $|\DeltaE|<50\mev$. Black points with error bars are the data, solid blue curves are the total PDF, long dashed red curves are the signal, dashed green curves are the continuum background, dotted magenta curves are the combinatorial $\BB$ background, and dash-dotted cyan curves are the feed-across background.}
\label{fig:2D}
\end{figure}

We determine the branching fraction as
\begin{equation}
 \mathcal{B}(B^{+} \rightarrow \KS \KS h^{+}) = \dfrac{n_{\mathrm{sig}}}{\epsilon  \times N_{\BB} \times [\mathcal{B}(\KS \rightarrow \pip \pim)]^{2}}\mbox{,}  
 \end{equation}
 where $n_{\mathrm{sig}}$, $\epsilon$, and $N_{\BB}$ are the total signal yield, average detection efficiency, and number of $\BB$ pairs, respectively.
Figure~\ref{fig:2D} shows signal enhanced $\DeltaE$ and $\nbprim$ projections of the separate fit to $B^{+}$ and $B^{-}$ samples for $\Bp\to \KS\KS\Kp$ and of the charge-combined fit for $\Bp\to \KS\KS\pip$.  For $B^{+}\to\KS\KS\pi^{+}$, we fit a total of 5103 candidate events to obtain a branching fraction of 
\begin{equation}
\mathcal{B}(B^{+} \rightarrow \KS \KS \pip)=(6.5 \pm 2.6 \pm 0.4)\times 10^{-7},
\end{equation} where the first uncertainty is statistical and the second is systematic (described below). Its signal significance is estimated as $\sqrt{-2\ln({\cal L}_0/{\cal L}_{\rm max})}$, where ${\cal L}_0$ and ${\cal L}_{\rm max}$ are the likelihood values for the fit with the branching fraction fixed to zero and for the best-fit case, respectively. Including systematic uncertainties by convolving the likelihood with a Gaussian function of width equal to the systematic uncertainty, we determine the significance to be $2.5$ standard deviations. In view of the significance being less than 3 standard deviations, we set an upper limit on the branching fraction of $\Bp\to \KS\KS\pip$. We integrate the convolved likelihood over the branching fraction to obtain the upper limit of $8.7 \times 10^{-7}$ at 90\% confidence level. This limit is similar to that of BaBar~\cite{BaBar:paper2}.

\begin{table*}[htpb]
\caption{Efficiency, differential branching fraction, and $\ACP$ in each $\mkk$ bin for $\Bp\to\KS\KS \Kp$.}
\label{tab:binfit}
\begin{center}
\begin{tabular}{c c c c c c c}
\hline\hline

 $\mkk$ (GeV/$c^{2}$) ~ & ~Efficiency (\%)~  & ~ & ~ $d{\cal B}/dM \times 10^{-6}$ ($c^{2}$/GeV)~ & ~&~ $\ACP$ (\%) ~ \\ 
  
 \hline 
 
  $1.0-1.1$  &    $24.0 \pm 0.4$& &$10.40 \pm 1.24 \pm 0.38$& & $~$ $-3.9 \pm 10.9 \pm 0.9$ \\

  $1.1-1.3$ &  $23.4 \pm 0.2$ & & $~$ $8.60 \pm 0.85 \pm 0.32$ & & $~~$ $-0.1 \pm 9.3 \pm 0.9$  \\
   
  $1.3-1.6$ &  $22.9 \pm 0.1$ & &$10.23 \pm 0.73 \pm 0.38$ & & $~~$ $+6.6 \pm 6.9 \pm 0.9$  \\

  $1.6-2.0$ &  $21.8 \pm 0.1$ & & $~$ $3.93 \pm 0.43 \pm 0.15$& &$+16.1 \pm 10.3 \pm 0.9$  \\
 
  $2.0-2.3$ &  $24.1 \pm 0.1$& & $~$ $3.90 \pm 0.47 \pm 0.15$ & & $~$ $-3.3 \pm 11.3 \pm 0.9$  \\
  
  $2.3-2.7$ &  $25.2 \pm 0.1$ & & $~$ $2.45 \pm 0.33 \pm 0.09$ & & $~$ $-5.7 \pm 12.2 \pm 1.0$  \\
  
  $2.7-5.0$ &  $26.3 \pm 0.0$ & &$~$ $0.35 \pm 0.07 \pm 0.01$ & & $-31.9 \pm 19.7 \pm 1.2$\\
   
\hline
\end{tabular}
\end{center}
\end{table*}

\begin{figure}[htpb]
\begin{center}
\includegraphics[scale=0.21]{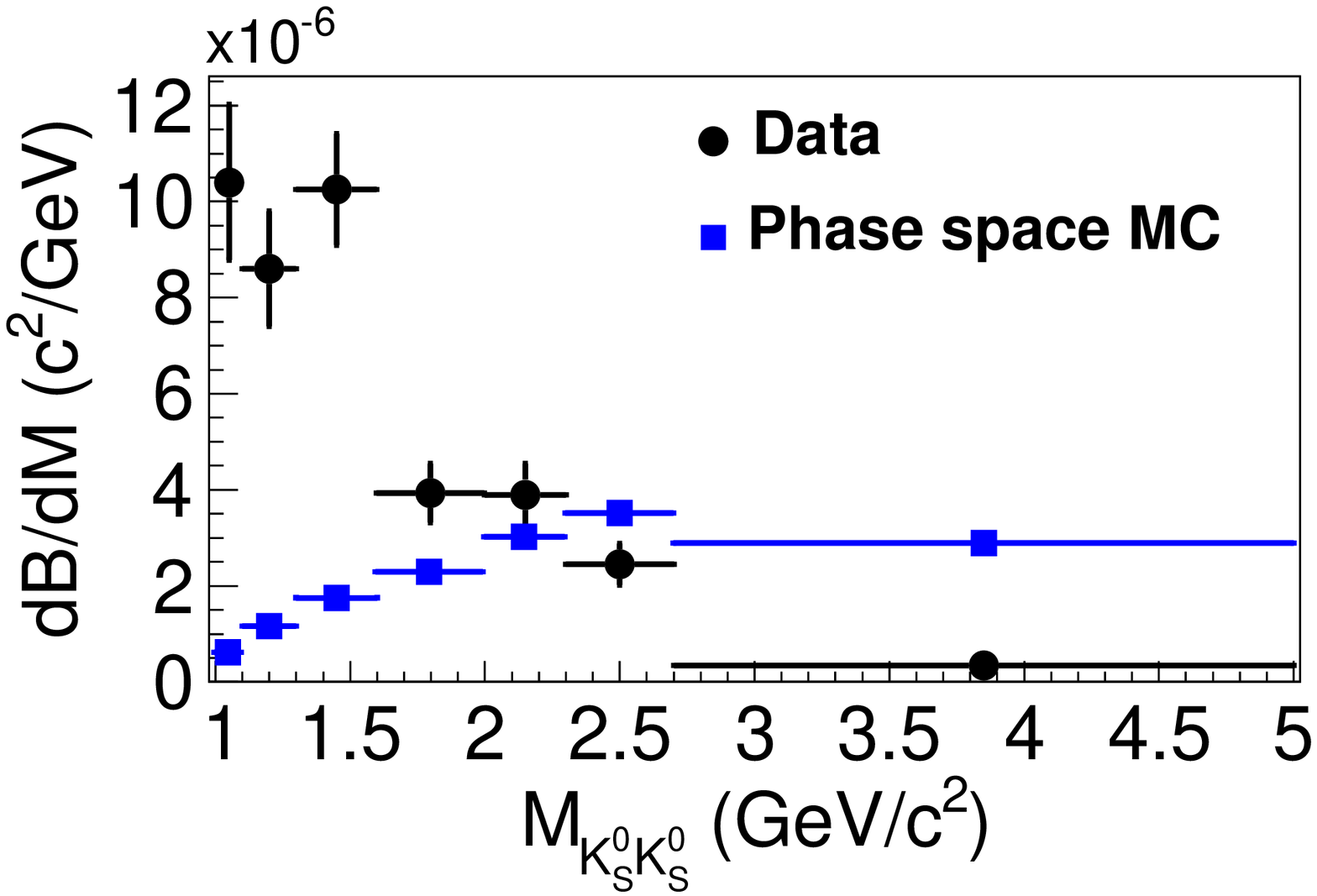} 
\includegraphics[scale=0.20]{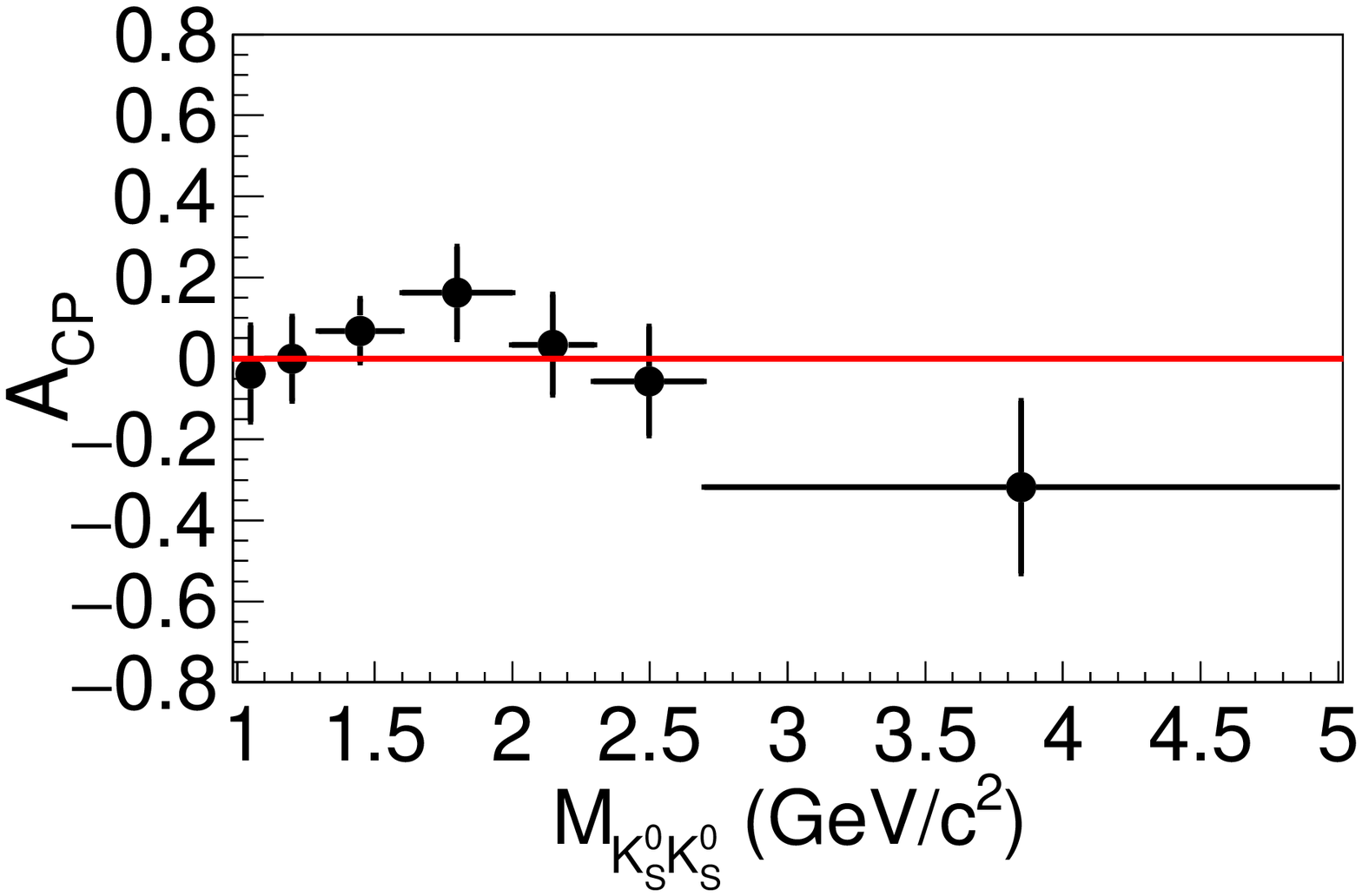} \\
\caption{Differential branching fraction (left) and $\ACP$ (right) as functions of $\mkk$ for $B^{+}\to\KS\KS K^{+}$. Black points with error bars are the results from the two-dimensional fits to data and include  systematic uncertainties. Blue squares in the left plot show the expectation from a phase space MC sample and the red line in the right plot indicates a zero $\CP$ asymmetry.}
\label{fig:binfit1}
\end{center}
\end{figure}

For $B^{+}\to\KS\KS K^{+}$, we perform the fit for 2709 candidate events in seven unequal bins of $\mkk$ to decipher contributions from possible quasi-two-body resonances. The efficiency, differential branching fraction, and $\ACP$ thus obtained  are listed in Table~\ref{tab:binfit}. Figure~\ref{fig:binfit1} shows the differential branching fraction and $\ACP$ plotted as a function of $\mkk$. We observe an excess of events around $1.5 \gevcc$ beyond the expectation of a 
phase space MC sample. No significant evidence for $\CP$ asymmetry is found in any of the bins. Upon inspection, no peaking structure beyond kinematic reflection is seen in the $\mkkp$ distribution. We calculate the  branching fraction by integrating the differential branching fraction over the entire $\mkk$ range: 
\begin{equation}
\mathcal{B}(B^{+} \rightarrow \KS \KS K^{+})=(10.42 \pm 0.43 \pm 0.22)\times 10^{-6}\mbox{,}
\end{equation}
where the first uncertainty is statistical and the second  is systematic. The $\ACP$ over the full $\mkk$ range is
\begin{equation}
\ACP(\Bp \rightarrow \KS \KS \Kp) = (+1.6 \pm 3.9 \pm 0.9)\% \mbox{.}
\end{equation}
This is obtained by weighting the $\ACP$ value in each bin with the obtained branching fraction in that bin. As the statistical uncertainties are bin-independent, their total contribution is a quadratic sum. For the systematic uncertainties, the contributions from the bin-correlated sources are linearly added, and those from the bin-uncorrelated sources are added in quadrature. The results agree with BaBar~\cite{BaBar:paper1}, which reported an $\ACP$ consistent with zero as well as the presence of quasi-two-body resonances $f_{0}(980)$, $f_0(1500)$, and $f'_2(1525)$ in the low $\mkk$ region.

Major sources of systematic uncertainty in the branching fractions are similar for both  $B^{+}\to\KS\KS K^{+}$ and $\KS\KS \pi^{+}$ decays. These are listed along with their contributions in Tables~\ref{tab:sys1} and~\ref{tab:sysm}. We use partially reconstructed $\Dstarp\to\Dz\pip$ with $\Dz \to\KS\pip\pim$ decays to assign the systematic uncertainty due to charged-track reconstruction ($0.35\%$ per track). The $\Dstarp\to\Dz\pip$ with $\Dz \to\Km\pip$ sample is used to determine the systematic uncertainty due to particle identification. The uncertainty due to the number of $\BB$ pairs is 1.37\%. The uncertainties due to continuum suppression and $M_{\mathrm{bc}}$ requirements are estimated with the control sample of $B^+\rightarrow \Dbar^0\pi^+$ with $\Dbar^0 \to\KS\pi^-\pi^+$. The uncertainty arising due to $\KS$ reconstruction is estimated from $D^{0}\rightarrow \KS \KS$ decays~\cite{Dash:paper}. A potential fit bias is checked by performing an ensemble test comprising $1000$ pseudoexperiments in which signal events are drawn from the corresponding MC sample and background events are generated according to their PDF shapes. The uncertainties due to signal PDF shape are estimated by varying  the correction factors by $\pm 1 \sigma$ of their statistical uncertainty. Similarly, the uncertainties due to background PDF shape are calculated by varying all fixed parameters by $\pm1\sigma$. We evaluate the uncertainty due to fixed background yields by varying them up and down by $20\%$ of their MC values.  The uncertainty due to fixed background $\ACP$ is estimated by varying the $\ACP$ values up and down by one unit of their statistical uncertainties. As for a possible systematics due to efficiency variation across the Dalitz plot in the $\Bp \to\KS\KS \pip$ channel, we find its impact to be negligible.
\begin{table}[htpb]
\caption{\label{tab:sys1}Systematic uncertainties in the branching fraction of $\Bp \to\KS\KS \pip$.}
\begin{center}
\begin{tabular}{lcc}
\hline\hline
{Source} &  Relative uncertainty in ${\cal B}$ ($\%$) \\
\hline
Tracking & $0.35$ \\
Particle identification & $0.80$ \\
Number of $\BB$ pairs & $1.37$ \\
Continuum suppression & $0.34$ \\
Requirement on $\mbc$ & $0.03$ \\
$\KS$ reconstruction & $3.22$ \\
Fit bias & $1.86$ \\
Signal PDF & $1.30$ \\
Combinatorial $\BB$ PDF & $+1.31, -1.98$ \\
Feed-across PDF& $+3.57, -4.10$ \\
Fixed background yield  & $+2.63, -2.27$  \\
Fixed background $ \ACP$ & $0.50$   \\
\hline
Total & $+6.30$, $-6.67$\\
\hline
\end{tabular}
\end{center}
\end{table}
\begin{table*}[htp]
\caption{\label{tab:sysm}Systematic uncertainties in the differential branching fraction and $\ACP$ in $\mkk$ bins for $\Bp \to\KS\KS \Kp$. ``$\dagger$'' indicates the uncertainty is independent of $\mkk$ with the listed value being applicable for all the bins. An ellipsis indicates a value below $0.05\%$ in $d{\cal B}/dM$ and 0.001 in~$\ACP$.}
\begin{center}
\begin{tabular}{lccccccc}
\hline\hline
 $M_{K^{0}_{S}K^{0}_{S}}$ (GeV/$c^2$) &  \begin{scriptsize} $1.0-1.1$ \end{scriptsize} & \begin{scriptsize}$1.1-1.3$   \end{scriptsize}  &  \begin{scriptsize}$1.3-1.6$ \end{scriptsize}   &  \begin{scriptsize} $1.6-2.0$  \end{scriptsize}  &  \begin{scriptsize} $2.0-2.3$  \end{scriptsize} &  \begin{scriptsize} $2.3-2.7$  \end{scriptsize} &  \begin{scriptsize} $2.7-5.0$ \end{scriptsize} \\

{Source} &  \multicolumn{5}{c}{Relative uncertainty in $d{\cal B}/dM$ ($\%$)}\\
\hline
Tracking$^\dagger$ & \multicolumn{7}{c}{$0.35$} \\
Particle identification$^\dagger$ & \multicolumn{7}{c}{$0.80$} \\
Number of $\BB$ pairs$^\dagger$ & \multicolumn{7}{c}{$1.37$} \\
Continuum suppression$^\dagger$ & \multicolumn{7}{c}{$0.34$} \\
Requirement on $M_{\mathrm{bc}}^\dagger$ & \multicolumn{7}{c}{$0.03$} \\
$\KS$ reconstruction$^\dagger$ & \multicolumn{7}{c}{$3.22$} \\
Fit bias$^\dagger$ & \multicolumn{7}{c}{$0.53$} \\
Signal PDF &  $^{+0.33}_{-0.27}$  & $^{+0.63}_{-0.48}$  & $^{+0.46}_{-0.44}$ & $^{+0.22}_{-0.63}$ & $^{+0.52}_{-0.38}$ & $0.67$ & $1.10$ \\
Combinatorial $\BB$ PDF &  $0.09$ & $^{+0.08}_{-0.13}$  & $0.12$ & $^{+0.17}_{-0.21}$ & $^{+0.26}_{-0.34}$ & $0.40$ & $0.40$ \\
Feed-across PDF &  $\cdots$ & $\cdots$  & $\cdots$ & $\cdots$ & $\cdots$ & $\cdots$ & $\cdots$ \\
Fixed background yield&  $\cdots$  & $0.10$&  $0.10$ & $0.23$ & $\cdots$ & $0.11$& $0.60$ \\
Fixed background $\ACP$&  $\cdots$  & $\cdots$&  $\cdots$ & $0.20$ & $0.10$ & $\cdots$& $0.13$ \\
\hline 
Total & $\pm 3.68$  & $\pm 3.72$ &  $\pm 3.69$ & $\pm 3.73$ & $\pm 3.72$ & $\pm 3.75$ & $\pm 3.89$\\
\hline
\multicolumn{2}{c}{} \\
\hline\hline
$M_{K^{0}_{S}K^{0}_{S}}$(GeV/$c^2$) & \begin{scriptsize} $1.0-1.1$  \end{scriptsize} & \begin{scriptsize}$1.1-1.3$   \end{scriptsize}  &  \begin{scriptsize}$1.3-1.6$ \end{scriptsize}   &  \begin{scriptsize} $1.6-2.0$  \end{scriptsize}  &  \begin{scriptsize} $2.0-2.3$  \end{scriptsize} &  \begin{scriptsize} $2.3-2.7$  \end{scriptsize} &  \begin{scriptsize} $2.7-5.0$ \end{scriptsize} \\  
Source & \multicolumn{5}{c}{Absolute uncertainty in $\ACP$} \\
\hline
Signal PDF &$0.001$ & $0.002$ & $0.001$&$0.002$ & $0.001$  & $0.001$& $0.004$ \\
Combinatorial $\BB$ PDF & $0.001$ & $0.001$ & $0.001$&$\cdots$ & $0.001$  & $0.002$& $0.001$ \\
Feed-across PDF &  $\cdots$ & $\cdots$  & $\cdots$ & $\cdots$ & $\cdots$ & $\cdots$ & $\cdots$ \\
Fixed background yield & $\cdots$ & $\cdots$& $0.001$& $0.001$& $0.001$& $0.001$  & $0.004$ \\
Fixed background $\ACP$  & $\cdots$ & $\cdots$& $0.001$& $0.001$& $0.001$ & $0.002$  & $0.006$ \\ 
Detector bias$^\dagger$ &  \multicolumn{7}{c}{$0.009$} \\
\hline
Total & $\pm 0.009$  & $\pm 0.009$ &  $\pm 0.009$ & $\pm 0.009$ & $\pm 0.009$ & $\pm 0.010$ & $\pm 0.012$\\
\hline
\end{tabular}
\end{center}
\end{table*}

Systematic uncertainties in $\ACP$ are listed in Table ~\ref{tab:sysm}. The systematic uncertainties due to the PDF  modeling, fixed background yields and $\ACP$ are estimated with the same procedure as for the branching fraction. Uncertainties due to the intrinsic detector bias on charged particle detection are evaluated with the samples of {$D^{+} \to \phi \pi^{+}$ and $D_{s}^{+} \to \phi \pip$ in conjunction with $D^{0} \to \Km \pip$~\cite{Dphipi:paper}. The total systematic uncertainty is calculated by summing all individual contributions in quadrature.

In summary, we have reported measurements of the charmless three-body decays $\Bp\to\KS\KS\Kp$ and $\Bp\to\KS\KS\pip$ using the full $\Y4S$ data sample collected with the Belle detector. We perform a two-dimensional simultaneous fit to extract the signal yields of both decays. For $\Bp\to\KS\KS\pip$, a 90\% confidence-level upper limit is set on the branching fraction at 8.7$\times$ $10^{-7}$. We measure the  branching fraction and $\ACP$ of $\Bp\to\KS\KS\Kp$ to be  
${\cal B}(\Bp\to\KS\KS\Kp) = (10.42\pm0.43\pm0.22)\times 10^{-6}$ and $\ACP(\Bp\to\KS\KS\Kp) = (+1.6\pm3.9\pm0.9)\%$. These results supersede Belle's earlier measurements~\cite{Belle:paper1} and are consistent with those of BaBar~\cite{BaBar:paper1,BaBar:paper2}.\\

We thank the KEKB group for the excellent operation of the
accelerator; the KEK cryogenics group for the efficient
operation of the solenoid; and the KEK computer group, and the Pacific Northwest National
Laboratory (PNNL) Environmental Molecular Sciences Laboratory (EMSL)
computing group for strong computing support; and the National
Institute of Informatics, and Science Information NETwork 5 (SINET5) for
valuable network support.  We acknowledge support from
the Ministry of Education, Culture, Sports, Science, and
Technology (MEXT) of Japan, the Japan Society for the 
Promotion of Science (JSPS), and the Tau-Lepton Physics 
Research Center of Nagoya University; 
the Australian Research Council including grants
DP180102629, 
DP170102389, 
DP170102204, 
DP150103061, 
FT130100303; 
Austrian Science Fund under Grant No.~P 26794-N20;
the National Natural Science Foundation of China under Contracts
No.~11435013,  
No.~11475187,  
No.~11521505,  
No.~11575017,  
No.~11675166,  
No.~11705209;  
Key Research Program of Frontier Sciences, Chinese Academy of Sciences (CAS), Grant No.~QYZDJ-SSW-SLH011; 
the  CAS Center for Excellence in Particle Physics (CCEPP); 
the Shanghai Pujiang Program under Grant No.~18PJ1401000;  
the Ministry of Education, Youth and Sports of the Czech
Republic under Contract No.~LTT17020;
the Carl Zeiss Foundation, the Deutsche Forschungsgemeinschaft, the
Excellence Cluster Universe, and the VolkswagenStiftung;
the Department of Science and Technology of India; 
the Istituto Nazionale di Fisica Nucleare of Italy; 
National Research Foundation (NRF) of Korea Grants
No.~2015H1A2A1033649, No.~2016R1D1A1B01010135, No.~2016K1A3A7A09005
603, No.~2016R1D1A1B02012900, No.~2018R1A2B3003 643,
No.~2018R1A6A1A06024970, No.~2018R1D1 A1B07047294; Radiation Science Research Institute, Foreign Large-size Research Facility Application Supporting project, the Global Science Experimental Data Hub Center of the Korea Institute of Science and Technology Information and KREONET/GLORIAD;
the Polish Ministry of Science and Higher Education and 
the National Science Center;
the Grant of the Russian Federation Government, Agreement No.~14.W03.31.0026; 
the Slovenian Research Agency;
Ikerbasque, Basque Foundation for Science, Spain;
the Swiss National Science Foundation; 
the Ministry of Education and the Ministry of Science and Technology of Taiwan;
and the United States Department of Energy and the National Science Foundation.

\end{document}

%% file: author-pub530.tex
\affiliation{University of the Basque Country UPV/EHU, 48080 Bilbao}
\affiliation{Beihang University, Beijing 100191}
\affiliation{University of Bonn, 53115 Bonn}
\affiliation{Brookhaven National Laboratory, Upton, New York 11973}
\affiliation{Budker Institute of Nuclear Physics SB RAS, Novosibirsk 630090}
\affiliation{Faculty of Mathematics and Physics, Charles University, 121 16 Prague}
\affiliation{Chonnam National University, Kwangju 660-701}
\affiliation{University of Cincinnati, Cincinnati, Ohio 45221}
\affiliation{Deutsches Elektronen--Synchrotron, 22607 Hamburg}
\affiliation{Duke University, Durham, North Carolina 27708}
\affiliation{University of Florida, Gainesville, Florida 32611}
\affiliation{Key Laboratory of Nuclear Physics and Ion-beam Application (MOE) and Institute of Modern Physics, Fudan University, Shanghai 200443}
\affiliation{Gifu University, Gifu 501-1193}
\affiliation{II. Physikalisches Institut, Georg-August-Universit\"at G\"ottingen, 37073 G\"ottingen}
\affiliation{SOKENDAI (The Graduate University for Advanced Studies), Hayama 240-0193}
\affiliation{Gyeongsang National University, Chinju 660-701}
\affiliation{Hanyang University, Seoul 133-791}
\affiliation{University of Hawaii, Honolulu, Hawaii 96822}
\affiliation{High Energy Accelerator Research Organization (KEK), Tsukuba 305-0801}
\affiliation{J-PARC Branch, KEK Theory Center, High Energy Accelerator Research Organization (KEK), Tsukuba 305-0801}
\affiliation{Forschungszentrum J\"{u}lich, 52425 J\"{u}lich}
\affiliation{IKERBASQUE, Basque Foundation for Science, 48013 Bilbao}
\affiliation{Indian Institute of Science Education and Research Mohali, SAS Nagar, 140306}
\affiliation{Indian Institute of Technology Bhubaneswar, Satya Nagar 751007}
\affiliation{Indian Institute of Technology Guwahati, Assam 781039}
\affiliation{Indian Institute of Technology Hyderabad, Telangana 502285}
\affiliation{Indian Institute of Technology Madras, Chennai 600036}
\affiliation{Indiana University, Bloomington, Indiana 47408}
\affiliation{Institute of High Energy Physics, Chinese Academy of Sciences, Beijing 100049}
\affiliation{Institute of High Energy Physics, Vienna 1050}
\affiliation{INFN - Sezione di Napoli, 80126 Napoli}
\affiliation{Advanced Science Research Center, Japan Atomic Energy Agency, Naka 319-1195}
\affiliation{J. Stefan Institute, 1000 Ljubljana}
\affiliation{Institut f\"ur Experimentelle Teilchenphysik, Karlsruher Institut f\"ur Technologie, 76131 Karlsruhe}
\affiliation{Kennesaw State University, Kennesaw, Georgia 30144}
\affiliation{King Abdulaziz City for Science and Technology, Riyadh 11442}
\affiliation{Department of Physics, Faculty of Science, King Abdulaziz University, Jeddah 21589}
\affiliation{Kitasato University, Sagamihara 252-0373}
\affiliation{Korea Institute of Science and Technology Information, Daejeon 305-806}
\affiliation{Korea University, Seoul 136-713}
\affiliation{Kyungpook National University, Daegu 702-701}
\affiliation{LAL, Univ. Paris-Sud, CNRS/IN2P3, Universit\'{e} Paris-Saclay, Orsay}
\affiliation{\'Ecole Polytechnique F\'ed\'erale de Lausanne (EPFL), Lausanne 1015}
\affiliation{P.N. Lebedev Physical Institute of the Russian Academy of Sciences, Moscow 119991}
\affiliation{Liaoning Normal University, Dalian 116029}
\affiliation{Faculty of Mathematics and Physics, University of Ljubljana, 1000 Ljubljana}
\affiliation{Ludwig Maximilians University, 80539 Munich}
\affiliation{Luther College, Decorah, Iowa 52101}
\affiliation{University of Malaya, 50603 Kuala Lumpur}
\affiliation{University of Maribor, 2000 Maribor}
\affiliation{Max-Planck-Institut f\"ur Physik, 80805 M\"unchen}
\affiliation{School of Physics, University of Melbourne, Victoria 3010}
\affiliation{University of Mississippi, University, Mississippi 38677}
\affiliation{University of Miyazaki, Miyazaki 889-2192}
\affiliation{Moscow Physical Engineering Institute, Moscow 115409}
\affiliation{Moscow Institute of Physics and Technology, Moscow Region 141700}
\affiliation{Graduate School of Science, Nagoya University, Nagoya 464-8602}
\affiliation{Kobayashi-Maskawa Institute, Nagoya University, Nagoya 464-8602}
\affiliation{Universit\`{a} di Napoli Federico II, 80055 Napoli}
\affiliation{Nara Women's University, Nara 630-8506}
\affiliation{National Central University, Chung-li 32054}
\affiliation{National United University, Miao Li 36003}
\affiliation{Department of Physics, National Taiwan University, Taipei 10617}
\affiliation{H. Niewodniczanski Institute of Nuclear Physics, Krakow 31-342}
\affiliation{Nippon Dental University, Niigata 951-8580}
\affiliation{Niigata University, Niigata 950-2181}
\affiliation{Novosibirsk State University, Novosibirsk 630090}
\affiliation{Osaka City University, Osaka 558-8585}
\affiliation{Pacific Northwest National Laboratory, Richland, Washington 99352}
\affiliation{Panjab University, Chandigarh 160014}
\affiliation{Peking University, Beijing 100871}
\affiliation{University of Pittsburgh, Pittsburgh, Pennsylvania 15260}
\affiliation{Punjab Agricultural University, Ludhiana 141004}
\affiliation{Theoretical Research Division, Nishina Center, RIKEN, Saitama 351-0198}
\affiliation{University of Science and Technology of China, Hefei 230026}
\affiliation{Seoul National University, Seoul 151-742}
\affiliation{Showa Pharmaceutical University, Tokyo 194-8543}
\affiliation{Soongsil University, Seoul 156-743}
\affiliation{Sungkyunkwan University, Suwon 440-746}
\affiliation{School of Physics, University of Sydney, New South Wales 2006}
\affiliation{Department of Physics, Faculty of Science, University of Tabuk, Tabuk 71451}
\affiliation{Tata Institute of Fundamental Research, Mumbai 400005}
\affiliation{Department of Physics, Technische Universit\"at M\"unchen, 85748 Garching}
\affiliation{Toho University, Funabashi 274-8510}
\affiliation{Department of Physics, Tohoku University, Sendai 980-8578}
\affiliation{Department of Physics, University of Tokyo, Tokyo 113-0033}
\affiliation{Tokyo Institute of Technology, Tokyo 152-8550}
\affiliation{Tokyo Metropolitan University, Tokyo 192-0397}
\affiliation{Utkal University, Bhubaneswar 751004}
\affiliation{Virginia Polytechnic Institute and State University, Blacksburg, Virginia 24061}
\affiliation{Wayne State University, Detroit, Michigan 48202}
\affiliation{Yamagata University, Yamagata 990-8560}
\affiliation{Yonsei University, Seoul 120-749}

 \author{A.~B.~Kaliyar}\affiliation{Indian Institute of Technology Madras, Chennai 600036} 
 \author{P.~Behera}\affiliation{Indian Institute of Technology Madras, Chennai 600036} 
 \author{G.~B.~Mohanty}\affiliation{Tata Institute of Fundamental Research, Mumbai 400005} 
 \author{V.~Gaur}\affiliation{Virginia Polytechnic Institute and State University, Blacksburg, Virginia 24061} 
 
  \author{I.~Adachi}\affiliation{High Energy Accelerator Research Organization (KEK), Tsukuba 305-0801}\affiliation{SOKENDAI (The Graduate University for Advanced Studies), Hayama 240-0193} 
  \author{J.~K.~Ahn}\affiliation{Korea University, Seoul 136-713} 
  \author{H.~Aihara}\affiliation{Department of Physics, University of Tokyo, Tokyo 113-0033} 
  \author{S.~Al~Said}\affiliation{Department of Physics, Faculty of Science, University of Tabuk, Tabuk 71451}\affiliation{Department of Physics, Faculty of Science, King Abdulaziz University, Jeddah 21589} 
  \author{D.~M.~Asner}\affiliation{Brookhaven National Laboratory, Upton, New York 11973} 
  \author{V.~Aulchenko}\affiliation{Budker Institute of Nuclear Physics SB RAS, Novosibirsk 630090}\affiliation{Novosibirsk State University, Novosibirsk 630090} 
% ...
 \collaboration {The Belle Collaboration}

%% file: paper_draft_final.bbl
\begin{thebibliography}{99}
\bibitem{charge:ref1}
Inclusion of charge-conjugate reactions are implicit unless stated otherwise.
\end{thebibliography}
